\documentclass[12pt]{article}
\usepackage{amsmath}
\usepackage{array}
\usepackage{graphicx}

\makeatletter \@addtoreset{equation}{section}

\makeatletter\renewcommand\section{\@startsection {section}{1}{\z@}%
                                   {-3.5ex \@plus -1ex \@minus -.2ex}
                                   {2.3ex \@plus.2ex}%
                                   {\normalfont\large\bfseries}}
\renewcommand\subsection{\@startsection{subsection}{2}{\z@}%
                                     {-3.25ex\@plus -1ex \@minus -.2ex}%
                                     {1.5ex \@plus .2ex}%
                                     {\normalfont\bfseries}}

\newcommand{\be}{\begin{equation}}
\newcommand{\ee}{\end{equation}}
\newcommand{\beq}{\begin{eqnarray}}
\newcommand{\eeq}{\end{eqnarray}}

\def\[{\left [}
\def\]{\right ]}
\def\({\left (}
\def\){\right )}

\def\CN{{\cal N}}

\def\r2{\sqrt{2}}

\def\CA{{\cal A}}

\def\CF{{\cal F}}
\def\CG{{\cal G}}

\def\CN{{\cal N}}

\newcommand{\bbibitem}[1]{\bibitem{#1}\marginpar{#1}}

\def\Label#1{\label{#1}%
  \smash{\hbox to0pt{\raise1ex\hbox{\tiny[#1]}\hss}}}
\def\noLabels{\let\Label=\label}
\def\nobbibitem{\let\bbibitem=\bibitem}

\begin{document}

\begin{titlepage}

\flushright{UCB-PTH-07/nn}
 \vfil\

\begin{center}

{\Large{\bf Constituent Model of Extremal non-BPS Black Holes\\ }}

\vspace{3mm}

Eric G. Gimon\footnote{e-mail: eggimon@lbl.gov}$^{\,a,b}$, Finn
Larsen\footnote{e-mail: larsenf@umich.edu}$^{\,c}$$^{,d}$ and Joan
Sim\'on\footnote{e-mail: J.Simon@ed.ac.uk}$^{\,e}$
\\

\vspace{8mm}

\bigskip\medskip
\smallskip\centerline{$^a$ \it Department of Physics, University of California, Berkeley,
CA 94720, USA.}
\medskip
\smallskip\centerline{$^a$ \it Physics Division, Lawrence Berkeley National Laboratory Berkeley, CA 94720, USA.}
\medskip
\smallskip\centerline{$^c$ \it Department of Physics, University of Michigan, Ann Arbor,
MI 48109, USA.}
\medskip
\smallskip\centerline{$^d$ \it Theory Division, CERN, CH-1211 Geneva 23, Switzerland.}
\medskip
\smallskip\centerline{$^e$ \it School of Mathematics and Maxwell Institute for Mathematical Sciences,}
\smallskip\centerline{\it King's Buildings, Edinburgh EH9 3JZ, Scotland.}

\vfil

\end{center}
\setcounter{footnote}{0}
\begin{abstract}
\noindent
We interpret extremal non-BPS black holes in four
dimensions as threshold bound states of four 1/2-BPS constituents. We verify
the no-force condition for each of the primitive constituents in the probe approximation.
Our computations are for a seed solution with $\overline{D0}-D4$ charges and 
equal $B$-fields, but symmetries extend the result to any U-dual frame.  
We make the constituent 
model for the $D0-D6$ system explicit, and also discuss a duality frame where the
constituents are $D3$ branes
at angles. 
We demonstrate stability of the constituent model in the weak coupling
description of the constituent D-branes. We discuss the relation between 
the BPS and non-BPS branches of configuration space. 
%
\end{abstract}
\vspace{0.5in}

\end{titlepage}
\renewcommand{\baselinestretch}{1.05}  

\newpage
\tableofcontents

\section{Introduction}

Non-BPS charged extremal black holes in four dimensions are interesting
because they represent an intermediate step between supersymmetric
extremal black holes and more physically realistic non-BPS non-extremal
uncharged black holes. In this paper we explore the physical properties of
non-BPS extremal black holes further with special emphasis on a constituent
model for them.

In N=2 supersymmetric theories with a U-duality action on the physical fields, some
properties of non-BPS extremal black holes can be obtained by analytical
continuation from their BPS relatives. This is the case for the black hole entropy,
perhaps the most prominent black hole characteristic. However, the black hole entropy, 
and analytical continuation generally,
is only part of the story. Other features of known explicit black hole solutions indicate
significant qualitative differences between the BPS and non-BPS branches of these U-duality
invariant theories, including:
\begin{enumerate}
\item
{\it Attractor Behavior:} The attractor mechanism for $N=2$ BPS black holes
applies to all scalars in vector multiplets but not those in hyper multiplets. 
For the symmetric $N=2$ theories which interest us
there are scalars in vector multiplets that decouple from the attractor flow so
that their horizon values are indeterminate. In other words, some scalars
experience a flat potential. It is remarkable
that it is the non-BPS black holes that exhibit the largest number of flat
directions in the supergravity approximation.
\item
{\it Mass Formula and Constituent Model:} The mass of extremal non-BPS black
hole in the supergravity limit can be written as the sum
of the masses of four primitive $1/2$ BPS constituents with no intrinsic
entropy of their own. This property applies everywhere in moduli space,
although the specific four-part split changes. The form of the mass formula
suggests that, at least in the supergravity limit, all non-BPS extremal black
holes are threshold bound states of four constituents. The analogous BPS 
mass formula is more involved so, again, there are certain remarkable
cancelations that apply specifically to non-BPS black holes in the supergravity
approximation. 
\item
{\it Phase Diagram:}
The mass of of the spherically symmetric non-BPS black holes is always strictly greater
than the BPS bound, even in regions of moduli space where BPS multi-center
solutions exist. This suggests that the two branches are related by a first order
phase transition.
\end{enumerate}

The starting point for this paper is the most general extremal static
spherically symmetric non-BPS black hole solution to the STU-model. This was first constructed
in \cite{gian}, and later rediscovered in a different U-duality framework in \cite{GLS07}, extending the work in \cite{Hotta:2007wz}\footnote{There is a lot of related work
in the literature concerning first order flow equations and attractors \cite{rwork} (see the review \cite{Ferrara:2008hwa} and references therein), microscopics \cite{Emparan:2006it,Emparan:2007en}, non-extremal D0-D6 interactions \cite{Camps:2008hb} and also quantum lift of flat directions \cite{Bellucci:2008tx}. The static single centered solutions studied here have also been recently extended to multi-center configurations with and without angular momentum \cite{Goldstein:2008fq,Bena:2009ev}.}.
The generating solution constructed in \cite{GLS07} takes a simple form
in a canonical duality frame where the charges are those of anti-D0-branes and
three kinds of $D4$-branes, while the three axionic scalars of the solution asymptote
a common $B$-field. In this canonical frame the mass of the extremal black
holes is :
\begin{equation}
  2G_N\,M_{\text{Non-BPS}}=\frac{1}{\sqrt{2}}
  \left(|Q_0|+\sum_{i=1}^3 P^i\,(1+B^2)\right)\,,
 \label{eq:nonBPSmass}
\end{equation}
with the convention that $Q_0<0$ on the non-BPS branch. 
The mass formula is simply the sum of the masses of anti-D0-branes and
D4-branes individually, with the $B$-field taken into account for each brane
independently. This suggests a constituent model with no binding energy, {\it
i.e.} a threshold bound state. In this paper we provide further evidence in
favor of this interpretation.  

To appreciate how surprising the non-BPS mass formula \eqref{eq:nonBPSmass}
is, let us compare with the
BPS mass formula :
\begin{equation}
  2G_N\,M_{\text{BPS}}=\frac{1}{\sqrt{2}}\left|Q_0+\sum_{i=1}^3 P^i\,(1+iB)^2\right|\,,
 \label{eq:BPSmass}
\end{equation}
where now $Q_0>0$. 
For non-vanishing $B$-field the total mass is less than the sum of constituent
masses. Hence we have a genuine bound state, with non-vanishing binding energy.
Although the binding energy vanishes in the limit where the $B$-field is
removed, it is believed that normalizable bound states persist in this limit
which corresponds to the BPS black hole.

Our constituent model of the non-BPS black holes as a threshold bound state
suggests a classical instability: throughout moduli
space we should be able to remove constituent quanta from the system and take
them to infinity, at no cost in energy. If we describe the effective dynamics of
such a constituent quantum as a probe in the background generated by the
extremal black hole, it must be the case that it feels no force, if it carries
the right charges to be interpreted as a constituent of the bound state. We
will test this expectation by explicit computation.

In a general U-duality frame, with arbitrary charge vector and asymptotic
moduli, the non-BPS mass formula similarly takes the form of a sum of four terms, each
of which is the mass of a 1/2-BPS constituent. 
Having identified the constituents of the non-BPS black hole 
in the canonical duality frame, the appropriate constituents for any other non-BPS black hole
can be determined as 
the image under U-duality of the canonical constituents. We will make this 
procedure explicit for the case of the $D0-D6$ system. 
In this case it is not obvious {\it a priori} what the 
four 1/2-BPS constituents should be. We find that the constituents of the
$D0-D6$ system are $D6$-branes with 
fluxes \cite{Taylor:1997ay,Larsen:1999pu} and verify that this gives the correct 
mass formula for the $D0-D6$ in the presence of general $B$-fields.  

We also consider a duality frame where the primitive constituents are interpreted
as $D3$-branes at angles. This more geometrical setting is well-suited for discussing
the spectrum of open strings stretching between the constituent branes. We will focus
on the stability condition imposed by the absence of tachyons. The representation of the
non-BPS black hole as $D3$-branes at angles is also well suited for discussing 
spacetime supersymmetry. 

Extremal non-BPS black hole have an instability into just two 1/2-BPS decay products, 
which are not mutually local. The $D0-D6$ frame realizes this instability in a simple manner:
overall energy is lowered if the $D0$-brane tunnels and escapes to infinity. 
Thus our threshold bound state is at best meta-stable. The full story is in fact more interesting
due to the existence of stable BPS solutions with lower energy than widely separated 
$D0$-brane and $D6$-branes. These
configurations necessarily have multiple centers, and they exist only when a sufficiently large 
$B$-field is turned on. Whenever there is a multi-center solution 
available, there may be a first order phase transition between the non-BPS
and the BPS branches.

This paper is organized as follows. In section 2 we review the canonical 
non-BPS solutions, with emphasis on the field strengths supporting the
solutions. In section 3 we present the probe computation verifying that
the proposed constituents feel no force from the non-BPS black hole, a
delicate matter in the presence of $B$-fields. 
In section 4 we determine the
constituent interpretation of the $D0-D6$ black hole. In this setting we
also discuss quantization conditions and the relation to BPS multicenter
solutions. In section 5 we examine a duality frame where the primitive constituents 
are $D3$-branes at angles. In this setting we discuss stability of the system from the world-sheet 
point of view, and we detail supersymmetry breaking.
Finally, we end in section 6 with a discussion of some open issues, including some comments
on the entropy of extremal non-BPS black holes. 
 
\section{The Canonical non-BPS Black Hole}

The setting for our study is the STU-model \cite{Cremmer:1984hj,Duff:1995sm,Behrndt:1996hu}, {\it i.e.} $N=2$ supergravity in four dimensions with
$n_V=3$ vector supermultiplets that couple through the prepotential :
\begin{equation}
  F =  \frac{s_{ijk} X^i\,X^j\,X^k}{6X^0} = \frac{X^1\,X^2\,X^3}{X^0}\,.
  \label{eq:prepot}
\end{equation}
The notation is $s_{ijk} = |\epsilon_{ijk}|$ with $i,j,k=1,2,3$.
The STU-model is a closed subsector of both $N=4$ and
$N=8$ supergravity and our results apply in those contexts as well, as detailed in \cite{GLS07}.

The single center extremal black hole solutions in the STU-model are uniquely
characterized by their charge vector $\Gamma = (P^I, Q_I)$ (with $I=0,1,2,3$) and the
asymptotic value of the complex moduli $z^i=X^i/X^0 = x^i - i y^i$ (with $i=1,2,3$). The STU-model has a $SL(2)^3$ duality
symmetry that acts nontrivially on these parameters so we may
consider a seed solution with just $(8+6)-9=5$ parameters with the understanding that
the most general charge vector and asymptotic moduli can be restored if needed, by acting with
dualities \cite{Cvetic:1996zq}.

Sufficient general black hole solutions are generally very complicated but there is
a canonical duality frame where the solution simplifies \cite{GLS07}. In this frame the five parameters
of the seed solution are four nonvanishing charges $Q_0$, $P^i$ and the
fifth parameter is chosen as the diagonal pseudoscalar $z^i = B-i$ (with the same $B$ for $i=1,2,3$).
We take $Q_0<0$ and $P^i>0$ which, in our conventions, means supersymmetry is
broken. With these choices, the four dimensional metric of the seed solution is :
\begin{equation}
  ds^2 = - e^{2U}\,dt^2 + e^{-2U} \left(dr^2 + r^2\,(d\theta^2 + \sin^2\theta\,d\phi^2)\right)\,,
  \label{eq:metric}
\end{equation}
with the conformal factor \cite{GLS07} :
\begin{equation}
  e^{-4U} = -4H_0H^1H^2H^3 - B^2\,,
\end{equation}
where the four harmonic functions are :
\begin{equation}
\sqrt{2}H_0 =
  -(1+B^2) + \frac{\sqrt{2}Q_0}{r}  \quad ,~\sqrt{2}H^i = 1+\frac{\sqrt{2}P^i}{r}~.
\end{equation}
The constants of integration have been adjusted so that the conformal factor
$e^{-4U}\to 1$ as $r\to\infty$. The conformal factor is positive definite because $Q_0<0$.
The scalar moduli $z^i$ are written in terms of the harmonic functions as
\begin{equation}
 z^i = \frac{B-i\,e^{-2U}}{s_{ijk}H^jH^k}\,.
 \label{eq:moduli}
 \end{equation}
 The asymptotic behavior is $z^i\to B-i$ as $r\to\infty$ in accord with the duality frame we
 have chosen.

The STU-model can be interpreted as a subsector of type IIA string theory on
$T^6 = T^2 \times T^2 \times T^2$. Then the scalars are the complexified K\"{a}hler
moduli $z^i = x^i - i y^i$ of the three $T^2$'s. The four electric charges correspond to
$D0$ and $D2$'s
wrapping the $T^2$'s, while the magnetic charges correspond to $D6$ and $D4$'s wrapping
the dual $T^4$'s.
Thus the canonical charge configuration that gives a simplified solution is the
$\overline{D0}$-D4 system, with identical $B$-field turned on in each of the $T^2$'s.

At this point we have not yet specified the gauge fields in the solutions. Since those
play a central role in the probe computations presented in the next section, it is
appropriate to derive them in detail. The result for the gauge fields is given in the end of this
section.

Starting from the gauge field $\vec{\CA}^I$ we introduce field strengths
$\CF^{\pm I}=d\CA^I\pm i\,\star d\CA^I$ $(I=1,2,3)$ that are imaginary anti-self-dual (imaginary self-dual) under
Hodge duality. The symplectic dual field strengths defined as :
\begin{equation}
  \CG_{\pm J} = \overline{\CN}_{JI}\,\CF^{\pm I}\,,
 \Label{eq:selfdual}
\end{equation}
are written analogously $\CG_{\pm J} = d\CA_J \pm i\,\star
d\CA_J$ in terms of the symplectic dual gauge field $\vec{\CA}_J$.
We decompose the field strength and the symplectic dual field strength into electric
and magnetic components as :
\begin{eqnarray}
  d\vec{\CA}^I &=& E^I\,dt\wedge dr + d\vec{a}^I\,,
   \Label{eq:formansatz}
 \\
  d\vec{\CA}_J &=& E_J\,dt\wedge dr + d\vec{a}_J\,.
   \Label{eq:formansat}
\end{eqnarray}
For single center solutions the Bianchi identities for $\CF^{I}$ and  $\CG_{J}$
determine the magnetic components uniquely in terms of conserved charges :
\begin{equation}
  \vec{a}^I = -P^I\,\cos\theta\,d\phi\,, \quad \vec{a}_J = -Q_J\,\cos\theta\,d\phi\,.
  \label{eq:magfields}
\end{equation}
The Bianchi identity for $\CG_{J}$ is equivalent to the equations of motion for the
"true" field strength $\CF^{I}$. The
magnetic charge of the symplectic dual field strength $\CG_{J}$ is
the electric charge in terms of  $\CF^{I}$ and therefore denoted $Q_J$.

At this point the gauge fields are completely specified but we must impose the
symplectic duality condition (\ref{eq:selfdual}) consistently in order to make them
explicit. Taking orientation so that $\epsilon_{{\hat t}{\hat r}{\hat \theta}{\hat \phi}}=+1$ the metric (\ref{eq:metric})
gives :
\begin{align}
  \star \left(dt\wedge dr\right) &= e^{-2U}\,r^2\,\left(d\theta\wedge \sin\theta\,d\phi\right)\,, \\
  \star \left(d\theta\wedge \sin\theta\,d\phi\right) &= - \frac{e^{2U}}{r^2}\,dt\wedge dr\,.
\end{align}
Note $\star^2 = -1$, as always on a four dimensional Lorentzian manifold.
The Hodge duals of the magnetic fields (\ref{eq:magfields}) become :
\begin{equation}
  \star d\vec{a}^I = -P^I\,\frac{e^{2U}}{r^2}\,dt\wedge dr\,, \quad
  \star d\vec{a}_J = -Q_J\,\frac{e^{2U}}{r^2}\,dt\wedge dr\,.
\end{equation}
Decomposing the moduli matrix $\overline{\CN}_{IJ}$ into
its real and imaginary part,
$\overline{\CN}_{IJ} = \mu_{IJ} - i\nu_{IJ}$
and focusing on the $dt\wedge dr$
components of the 2-forms in \eqref{eq:selfdual} we
solve for the electric components $\{E^I,\,E_J\}$ introduced
in (\ref{eq:formansatz}-\ref{eq:formansat}) and find :
\begin{align}
  E^I &= \frac{e^{2U}}{r^2}\,\nu^{IJ}\,\left[Q_J
  - \mu_{JK}P^K\right]\,,
  \Label{eq:E1} \\ E_J &= \frac{e^{2U}}{r^2}\,\left[\mu_{JK}
\nu^{KL}Q_L -\left(\mu_{JK}
\nu^{KL}
\mu_{LM}+\nu_{JM}\right) P^M\right]\,. \Label{eq:E2}
\end{align}
The matrix $\nu^{IJ}$ is the inverse of $\nu_{IJ}$, {\it i.e.}
$\nu^{IJ}\,\nu_{JK} = \delta^I\,_K$.  The $d\theta\wedge d\phi$
components of \eqref{eq:selfdual} give no further
constraints, they give equations that are satisfied automatically.

The expressions (\ref{eq:E1}-\ref{eq:E2}) are general, valid for $N=2$ supergravity with
any number of vector multiplets. However, they depend on the scalar fields through the moduli
matrix defined as :
  \begin{align}
  \CN_{IJ} &= \mu_{IJ} + i\nu_{IJ}= \overline{F}_{IJ} +
  2i\,\frac{ (\text{Im}\,F_{IK})X^K\,(\text{Im}\,F_{JM})X^M}{(\text{Im}\,F_{RS})X^RX^S}\,,
  \Label{eq:mmatrix} \\
  F_{IJ} &= \frac{\partial^2 F}{\partial X^I\partial X^J}\,,
\end{align}
and to make that dependence explicit we need the
prepotential, {\it i.e.} (\ref{eq:prepot}) in the case of the STU-model.
Writing out \eqref{eq:mmatrix} in this case we find \cite{Behrndt:1996hu} :
\begin{equation}
\mu_{IJ} =
\begin{pmatrix}
  2x_1x_2x_3 & -x_2x_3 & -x_1x_3 & -x_1x_2 \cr
  -x_2x_3 & 0 & x_3 & x_2 \cr
  -x_1x_3 & x_3 & 0 & x_1 \cr
  -x_1x_2 & x_2 & x_1 & 0 \cr
\end{pmatrix}\,,
\Label{eq:rmmatrix}
\end{equation}
from the real part of the equation and :
\begin{equation}
\nu_{IJ} =y_1y_2y_3
\begin{pmatrix}
  -\left(1 + \,\frac{x_1^2}{y^2_1} + \,\frac{x_2^2}{y^2_2}+\,\frac{x_3^2}{y^2_3}
 \right) &
  \frac{x_1}{y_1^2} & \frac{x_2}{y_2^2} &
  \frac{x_3}{y_3^2} \cr \frac{x_1}{y_1^2} & -
  \frac{1}{y_1^2} & 0 & 0 \cr \frac{x_2}{y_2^2} & 0 & -
  \frac{1}{y_2^2} & 0 \cr \frac{x_3}{y_3^2} & 0 & 0 & -
  \frac{1}{y_3^2}\cr
\end{pmatrix}\,,
\Label{eq:immatrix}
\end{equation}
from the imaginary part.
We also need the inverse of \eqref{eq:immatrix} :
\begin{equation}
  \nu^{IJ} = -\frac{1}{y_1y_2y_3}
  \begin{pmatrix}
    1 & x_1 & x_2 & x_3 \cr
    x_1 & x_1^2 + y_1^2 & x_1x_2 & x_1x_3 \cr
    x_2 & x_1x_2 & x_2^2+y_2^2 & x_2x_3 \cr
    x_3 & x_1x_3 & x_2x_3 & x_3^2 +y_3^2\cr
  \end{pmatrix}\,.
\Label{eq:invimmatrix}
\end{equation}
The position of the indices on the real moduli is usually taken lower ({\it i.e.} $x_i, y_i$)
for typographical convenience although, strictly, these fields have only been defined with
upper indices ($z^i = x^i - i y^i$).

The electric fields supporting the seed solution \eqref{eq:metric} is found from the general expressions \eqref{eq:E1}
by inserting the moduli matrices $\nu_{IJ}, \mu_{IJ}$ and turning on only the charges $(Q_0,\,P^i)$.
The result is :
\begin{eqnarray}
  E^0 &=& -\frac{e^{2U}}{r^2}\,\frac{1}{y_1y_2y_3}\,\left(Q_0-\frac{1}{2}s_{ijk}x^ix^jP^k\right)\,, \label{eq:es0} \\
  E^i &=& -\frac{e^{2U}}{r^2}\,\frac{1}{y_1y_2y_3}\,\left(x^i\,Q_0 - (x_i^2+y_i^2)\,s_{ijk} x_j\,P^k-\frac{1}{6}P^i\,s_{jkl}x^jx^kx^l\right)\,, \label{eq:esi}
\end{eqnarray}
where there is no summation over the free index $i$ in the expression for $E^i$.
The electric fields give the forces on electric probes. The full electromagnetic field is given in \eqref{eq:formansatz} with the electric fields
(\ref{eq:ess0}-\ref{eq:esi}) and the magnetic fields \eqref{eq:magfields}. We will also need the dual electric fields
\eqref{eq:E2} :
\begin{eqnarray}
  E_0 &=& -\frac{e^{2U}}{r^2}\,\frac{1}{y^1y^2y^3}\,\left(-x^1x^2x^3\,Q_0 + \frac{1}{2}x_iP^i\,s_{ijk}(x_j^2+y_j^2)(x_k^2+y_k^2)\right)\,, \label {eq:ess0} \\
  E_i &=& -\frac{e^{2U}}{r^2}\,\frac{s_{ijk}}{2y^1y^2y^3}\,\left(x_jx_k\,Q_0-(x_j^2+y_j^2)\,(x_k^2+y_k^2)\,P^i -2x_ix_j\,(x_k^2+y_k^2)\,P^j\right)\,. \label{eq:essi}
\end{eqnarray}
Again, there is no summation over the free index $i$ in the expression for $E_i$. The
dual electric fields give the forces on magnetic probes.

\section{Probing Extremal Non-BPS Black Holes}
As explained in the introduction, it is reasonable to describe non-BPS black
holes as threshold bound states of $1/2$-BPS constituents. The canonical
$\overline{D0}-D4$ solution reviewed in the previous section is thus
interpreted as a collection of $\overline{D0}$-brane and $D4$-brane
constituents placed on top of each other with no binding energy.

In this section we test the interpretation as follows. The lack of binding
energy means constituents can be arbitrarily separated. Thus it should be
possible to bring in additional constituents from infinity, without them being
subject to a force. Accordingly, we expect $\overline{D0}$-branes and
$D4$-branes wrapping any two torii to feel no force, whereas other
$1/2$ BPS probes like wrapped $D2$'s and $D6$'s should
feel forces.

The potential felt by a static $Dp$-brane at a constant position due to a
background field is given by the Lagrangian density of the $Dp$-brane, up to a
sign :
\begin{equation}
  V_{Dp} = T_p \left[ e^{-(\phi-\phi_\infty)}\,\sqrt{-\det (G+B)} - \sqrt{2}\eta\,A_{p+1}\right] \equiv T_p\left( V_{\text{DBI}} + V_{\text{WZ}}\right)\,,
  \label{eq:DBI}
\end{equation}
where $\eta$ parameterises whether we are describing a $Dp$ or an
$\overline{Dp}$ brane. We have in mind infinitesimal constituents being added and so
it is justified to use the probe approximation
where distortion of the background due to the probe is neglected. The DBI action should give
a precise description even though the background is non-BPS, because the proposed
constituents are $1/2$-BPS.

The dilaton in \eqref{eq:DBI} is the 10D dilaton, with its asymptotic value
absorbed in the tension of the brane. The 4D dilaton is a component of a hypermultiplet,
which has no radial dependence, so the 10D dilaton acquires its variation solely from
the volume of $T^6$ in the condition $e^{-2(\phi_4-\phi_{4\infty})}=e^{-2(\phi-\phi_\infty)}V_6=1$. It is convenient to evaluate the combination :
\begin{equation}
  e^{-(\phi-\phi_\infty)}\,\sqrt{-g_{tt}} = \frac{1}{\sqrt{y^1y^2y^3}} e^U =
  \frac{2\sqrt{2}H^1 H^2 H^3}{(-I_4 - B^2)}
  \,.
  \label{eq:effdil}
\end{equation}
Here we are introducing the conventional notation for the quartic duality invariant :
\be
I_4 = 4H_0H^1H^2H^3 - 4H^0H_1H_2H_3 - \(\sum_I H_I H^I \)^2
+ 4 \sum_{i<j} H^i H_i H^j H_j~,
\label{eq:i4def}
\ee
which the charge assignments of the seed solution reduces to :
\be
I_4 \to\;
4H_0H^1H^2H^3\,.
\ee
Recall that $I_4<0$ for non-BPS solutions.

We have written the DBI-action \eqref{eq:DBI} in the conventional manner but we should
remember that the $B$-field appearing in \eqref{eq:DBI} is the spatially varying $B$-field,
whose components on each $T^2$ we hitherto denoted $x$. For a single $T^2$ the dictionary
of notations is :
\begin{equation}
\sqrt{\det (G+B)}  \to \sqrt{x^2 + y^2} = |z|   \,.
\end{equation}

The normalization of the WZ-term in \eqref{eq:DBI} is unconventional, a consequence of
the definition of gauge fields we adopted\footnote{We have checked that the normalization
given here gives the correct BPS conditions. Also, for non-BPS states the coefficient is determined
by cancellation of forces in the absence of a $B$-field and then the cancellation for general
$B$-field is independent of conventions.}. The contribution to the force (in units of the brane tension $T_p$)
from the WZ term is simply:
\begin{equation}
- \frac{\partial V_{WZ}}{\partial r} = \sqrt{2}\eta \frac{\partial A_{p+1}}{\partial r}
= -\sqrt{2}\eta E\,.
\label{eq:WZforce}
\end{equation}
The electric field one should use in this expression depends on the identity of the probe:
it is $E^0$, $E^i$ for $D0$, $D2$ branes and $E_0$, $E_i$ for $D6$, $D4$ branes.
The electric fields generated by the $\overline{D0}$-D4 background
were given in (\ref{eq:es0}-\ref{eq:essi}).

We are now ready to compute the forces that the seed solution exerts on a variety of probes.
In the following we establish that the $\overline{D0}-D4$ background exert no forces on
$\overline{D0}$-branes, nor on $D4$-branes, despite the presence of a $B$-field. These
results support our contention that these are the constituents of the bound state. As a
means of emphasizing the nontrivial nature of the cancellations, we also carry out the
corresponding computation for a BPS black hole and show that, in that case, the
$B$-field obstructs the cancellation.

\subsection{D0-brane probe}
Consider a $D0$-brane (or a $\overline{D0}$-brane) experiencing the forces of the extremal non-BPS $\overline{D0}-D4$ black
hole with equal B fields. The DBI contribution to the force is found by differentiating \eqref{eq:effdil} :
\begin{eqnarray}
 -  \frac{\partial V_{\text{DBI}}}{\partial r} &=&
-  \frac{2\sqrt{2}H^1H^2H^3}{(-I_4-B^2)^2}\,\left( \frac{\partial I_4}{\partial r}
 + (-I_4 - B^2)\sum_i \frac{1}{H^i}\frac{\partial H^i}{\partial r}\right)
\cr
&=& \frac{1}{r^2}  \frac{2\sqrt{2}H^1H^2H^3}{(-I_4-B^2)^2}\,\left(I_4\,\frac{Q_0}{H_0}-
  B^2\sum_i \frac{P^i}{H^i}\right)\,.
  \label{eq:d0att}
\end{eqnarray}
Since $I_4<0$ the force is negative, {\it i.e.} towards smaller $r$, as one expects for the attractive
gravitational and dilatonic forces. The DBI-contribution to the force is the same for a $D0$-brane
and for a $\overline{D0}$-brane.

The WZ contribution to the force is given by inserting \eqref{eq:es0} in
\eqref{eq:WZforce} :
 \begin{eqnarray}
- \frac{\partial V_{WZ}}{\partial r} &=& -\sqrt{2}\eta E^0 \,. \cr
&=& \frac{\sqrt{2}\eta}{r^2} \frac{8(H^1H^2H^3)^2}{(-I_4-B^2)^2}\left( Q_0 - \frac{B^2}{4H^1 H^2 H^3}\sum_i \frac{P^i}{H^i}
 \right)
\cr
&=& \frac{\eta}{r^2}  \frac{2\sqrt{2}H^1H^2H^3}{(-I_4-B^2)^2}\,\left(I_4\,\frac{Q_0}{H_0}-
  B^2\sum_i \frac{P^i}{H^i}\right)\,.
\end{eqnarray}
In the second line we simplified using the second part of \eqref{eq:effdil}. The
result for the WZ-force is positive for $\eta=-1$ and negative for $\eta=+1$, because
$\overline{D0}$'s are repelled from the $\overline{D0}-D4$ black hole, while $D0$'s
are attracted. In the case of $\overline{D0}$ the repulsion precisely cancels the
attraction \eqref{eq:d0att} such that there is no net force, even in the presence of a $B$-field.
This result confirms our expectation that the extremal black hole contains  $\overline{D0}$
constituents at threshold.

\subsection{D4-brane probe}

We consider a D4-brane wrapping tori one and two (without losing
generality). From \eqref{eq:DBI} we find the potential felt by a static configuration :
\begin{eqnarray}
V_{\rm DBI}    &=& e^{-\phi}\,\sqrt{-g_{tt}}\,|z_1||z_2|
 \cr &=&
\frac{1}{\sqrt{2}H^3}\frac{-I_4}{(-I_4 - B^2) }\,,
\end{eqnarray}
where we used \eqref{eq:effdil} and wrote the moduli \eqref{eq:moduli} in the useful
form :
\begin{equation}
  |z_i|^2 = \frac{-I_4}{(s_{ijk}H^jH^k)^2}\,.
\end{equation}
The DBI contribution to the force then becomes :
\begin{equation}
 - \frac{\partial V_{\text{DBI}}}{\partial r} = \frac{1}{r^2}
 \frac{\,I_4}{\sqrt{2}H^3 (-I_4-B^2)^2}\left[ - I_4 \frac{P^3}{H^3} +
  B^2 \left(
  \frac{Q_0}{H_0}+\frac{Q_1}{H_1}+\frac{Q_2}{H_2}\right)\right]\,.
  \label{wzd4force}
\end{equation}
after some simplifications. Since $I_4<0$ the force is negative again, as one
expects for the attractive gravitational and dilatonic forces.

The WZ contribution to the force is given by inserting \eqref{eq:essi} in
\eqref{eq:WZforce} :
 \begin{eqnarray}
- \frac{\partial V_{WZ}}{\partial r} &=& -\sqrt{2}\eta E_3 \,. \cr
&=&
\frac{\eta}{r^2}
 \frac{\,I_4}{\sqrt{2}H^3 (-I_4-B^2)^2}\left[ - I_4 \frac{P^3}{H^3} +
  B^2 \left(
  \frac{Q_0}{H_0}+\frac{Q_1}{H_1}+\frac{Q_2}{H_2}\right)\right]\,
\end{eqnarray}
The force is positive for $\eta=-1$ so we interpret that sign as corresponding to a $D4$-brane,
the case where the probe is repelled from the $\overline{D0}-D4$ black hole
background. For $\eta=-1$ there is a perfect cancelation between DBI and WZ forces at
all positions, and with general $B$ taken into account. This supports our interpretation
of the $D4$ as one of the 1/2-BPS constituents of the non-BPS black hole.

\subsection{Other Probes}
Our formulae easily gives the forces on many other probe branes, such as $D2$, $D6$,
and also various branes with fluxes turned on.

The case of $D2$, $D6$ is particularly simple. In the absence of a $B$-field there is
just the attractive force due to the gravity-dilaton interactions encoded in the DBI
action and the WZ-term vanishes identically because the charges involved in
background and in probe are different. The inclusion of a $B$-field makes the accounting
less transparent, because the $B$-field induces electric fields of all types and so
a contribution from the WZ-term. Nevertheless, a net attractive force remains
even when $B$ is taken into account.

A more subtle case is when we consider $\overline{D0}$, $D4$ with fluxes on their
world-volumes. The fluxes modify the DBI term and also the WZ term, obstructing the
delicate cancellation exhibited above in the absence of fluxes. This gives rise to a net
force on the probe. This shows that the correct constituents for the $\overline{D0}-D4$
black hole are the $\overline{D0}$'s and $D4$'s with no fluxes on their
world-volumes.

\subsection{A Supersymmetric Probe Computation}
The cancellation of forces made explicit in the preceding subsections is reminiscent of
similar phenomena in simple supersymmetric systems. In order to appreciate that the
non-BPS cancellations we exhibit are in fact novel, it is worth carrying out analogous
computations for BPS black holes.

To do this let us consider the standard BPS black holes with $D0$, $D4$ charges
and a diagonal $B$-field \cite{Behrndt:1997ny,Denef:2000nb,Bates:2003vx}.
The metric remains of the form \eqref{eq:metric}
but in the BPS case the conformal factor is :
\begin{equation}
e^{-4U_{\rm BPS}} = I_4~,
\end{equation}
where the quartic invariant $I_4$ defined in \eqref{eq:i4def} depends on the harmonic functions :
\begin{eqnarray}
H^0 & = & \overline{P}^0\quad,~
H^i  = \overline{P} + \frac{P^i}{r}~, \cr
H_0 & = &  \overline{Q}_0 + \frac{Q_0}{r}\quad,~
H_i = \overline{Q}~.
\end{eqnarray}
We have $I_4>0$ for BPS configurations.
The $B$-field is encoded in the constants :
\begin{eqnarray}
\overline{P}^0 & = &  \frac{1}{\sqrt{2}}\sin\alpha\quad,~
\overline{Q}_0 = \frac{1}{\sqrt{2}}\left((1-3B^2) \cos\alpha + B(3-B^2)\sin\alpha\right)~,
\cr
\overline{P}  &=& \frac{1}{\sqrt{2}} (\cos\alpha + B\sin\alpha)
\quad,~
\overline{Q} =  \frac{1}{\sqrt{2}} \left( 2B\cos\alpha -\sin\alpha ( 1- B^2)\right)~,
\end{eqnarray}
where the phase of the spacetime central charge is :
\begin{eqnarray}
\tan\alpha = \frac{2B\sum_i P^i}{Q_0 + \sum_i P^i (1-B^2)}~.
\end{eqnarray}
The scalar fields in the BPS solution are :
\begin{equation}
z^i = \frac{(H^I H_I - 2H^i H_i) -ie^{-2U}}{s_{ijk} H^j H^k - 2H^0 H_i}~,
\end{equation}
with no sum over the index $i$.

Let us consider a $D0$-brane probe. In this case the DBI potential becomes :
\begin{eqnarray}
V_{\rm DBI} & = & \frac{1}{\sqrt{y^1 y^2 y^3}}e^U \cr
&=& \frac{\sqrt{8(H^2 H^3 - H^0 H_1)(H^3 H^1 - H^0 H_2)(H^1 H^2 - H^0 H_3)}}{I_4} \cr
&~& \left( 1 + 2\overline{P}\sum_i P^i\frac{1}{r} \right)
\left[ 1 - \sqrt{2} \left( (Q_0 + \sum_i P^i (1 - B^2)) \cos\alpha + 2B\sum_i P^i \sin\alpha\right)\frac{1}{r} \right]
\cr &\sim &
1 - \sqrt{2} \left( Q_0 \cos\alpha+ B\sum_i P^i (\sin\alpha - B\cos\alpha)  \right) \frac{1}{r}
\end{eqnarray}
We expanded for large $r$ using :
\begin{equation}
I_4 = 1 + \sqrt{2} \left( (Q_0 + \sum_i P^i (1 - B^2)) \cos\alpha + 2B\sum_i P^i \sin\alpha\right){1\over r} + \cdots~.
\end{equation}
We then find the gravity-dilaton force :
\begin{eqnarray}
- \frac{\partial V_{\rm DBI}}{\partial r} &= &
- \sqrt{2} \left( Q_0 \cos\alpha+ B\sum_i P^i (\sin\alpha - B\cos\alpha)  \right) \frac{1}{r^2} +\cdots
\end{eqnarray}

The WZ-coupling is written in terms of the electric field in \eqref{eq:WZforce} and
the applicable electric field is given in \eqref{eq:ess0}. This
gives the force :
\begin{eqnarray}
-  \frac{\partial V_{\rm WZ}}{\partial r} & =& -\sqrt{2}\eta E^0 \cr
& = & \eta \frac{\sqrt{2}}{r^2} {e^{2U}\over y_1 y_2 y_3} \left[ Q_0 - (x_2 x_3 P^1  + x_3 x_1 P^2+ x_1 x_2 P^3 )\right]\cr
& =  & \frac{\eta \sqrt{2}}{ r^2}(Q_0 - B^2 \sum_i P^i) + {\cal O}(\frac{1}{r^3})~.
\end{eqnarray}

The case of $\eta=1$ corresponds to a $D0$-brane (rather than an $\overline{D0}$-brane).
In this case the force cancels completely when there is no $B$-field, as one expect
for a BPS system, but generally the $B$-field obstructs the cancellation :
\begin{eqnarray}
- \frac{\partial V_{\rm WZ}}{\partial r} - \frac{\partial V_{\rm DBI}}{\partial r} &= &
\sqrt{2} \left[ Q_0 ( 1- \cos\alpha)  - B\sum_i P^i (B (1 - \cos\alpha)  +\sin\alpha )\right] \frac{1}{r^2}+ {\cal O}(\frac{1}{r^3})\cr
&  = & -{2B^2\over r^2} { (\sum_i P^i)^3\over (Q_0 + \sum_i P_i)^2}+ {\cal O}(\frac{1}{r^3})~.
\end{eqnarray}
Thus there is generally a net force in the supersymmetric system. The force is negative,
{\it i.e.}  attractive, indicating that the $D0$'s are bound to the $D0-D4$ system in the
presence of a $B$-field. That is indeed what we expect from the BPS mass formula \eqref{eq:BPSmass},
which indicates that there is a genuine bound state, {\it i.e.} one with binding energy.

The point we emphasize in this section is that the analogous non-BPS state is very different
from the BPS state: there is no force whatsoever even in the presence of a $B$-field.

\section{U-duality and the $D0-D6$ Black Holes}

The non-BPS black hole considered so far is a seed solution. This means any 
other extremal non-BPS black hole solution can be generated by acting with $U$-duality. 
Acting with $U$-duality on the four primitive constituents identified for the seed solution,
we can construct the primitive constituents appropriate for any extremal black hole we wish to 
analyze. By construction such primitive constituents will feel no forces from the black hole
in the probe approximation. This matches the U-duality invariance of the probe potential computed before \cite{Denef:2002ru}. Accordingly we interpret a general non-BPS black hole
as a marginal bound state of the corresponding four primitive constituents. 

In this section we use employ U-duality to analyze the non-BPS extremal $D0-D6$ black hole
in the presence of background $B$-fields. We find that the constituents are $D6$-branes with 
specific fluxes turned on. In the regime with large $B$-fields there exist BPS configurations with 
the same charges as the black holes we consider. We use this circumstance to clarify the relation 
between the BPS and the non-BPS branches. 

\subsection{$D0-D6$ constituents and the DBI Mass}
The five parameters of the seed solution ($\overline{D0}$-charge, three $D4$-charges, and 
a common $B$-field along three $T^2$'s) can be mapped by U-duality
to the five parameters of the $D0-D6$ black hole ($D0$-charge $Q_0$, 
$D6$-charge $P^0$, and independent $B$-fields $B_{1}, B_{2}, B_{3}$ along three $T^2$'s). 
The explicit map (constructed in section 5 of \cite{GLS07}) depends prominently 
on three parameters $\Lambda_i$ related to the variables of the $D0-D6$ frame by the equations 
\begin{eqnarray}
\Lambda_1\Lambda_2\Lambda_3 &=&  \frac{P^0}{Q_0}\,,\label{eq:prodlambda} \\
{1\over 2}[ \Lambda_1(1 + B_{1}^2) - \Lambda_1^{-1} ]&=& {1\over 2}[\Lambda_2(1 + B_{2}^2) -
\Lambda_2^{-1} ]= {1\over 2} [ \Lambda_3(1 + B_{3}^2) - \Lambda_3^{-1}] \,. 
\label{eq:sugcons} 
\end{eqnarray}
The awkward constraint \eqref{eq:sugcons} arises from the requirement that the $B$-field in the
$\overline{D0}-D4$ seed solution is the same on the three $T^2$'s. We included the factor of $1/2$ so that these
expressions are precisely dual to the $B$-field of the seed solution. 

The complete solution describing the $D0-D6$ black hole in the presence of $B$-fields follows by 
substituting the explicit duality map into the seed solution. The resulting expressions are unwieldy and 
not very illuminating, so we will not present them here. 
A more instructive computation is to transform the four primitive constituents of the seed solution by the
duality transformation and so identify the primitive constituents underlying the $D0-D6$ black hole. 
This transformation gives the charge vectors
\begin{eqnarray}
\label{constvector}
  \Gamma_{I} &=&
{1\over 4}  \Big(P^0;-P^0/\Lambda_1,-P^0/\Lambda_2,-P^0/\Lambda_3;Q_0;P^0/(\Lambda_2\Lambda_3),
P^0/(\Lambda_1\Lambda_3),P^0/(\Lambda_1\Lambda_2)\Big)\label{cvecone}\\
  \Gamma_{II} &=&
{1\over 4}  \Big(P^0;-P^0/\Lambda_1,P^0/\Lambda_2,P^0/\Lambda_3;Q_0;P^0/(\Lambda_2\Lambda_3),
-P^0/(\Lambda_1\Lambda_3),-P^0/(\Lambda_1\Lambda_2)\Big) \label{cvectwo}\\
  \Gamma_{III} &=&
{1\over 4}  \Big(P^0;P^0/\Lambda_1,-P^0/\Lambda_2,P^0/\Lambda_3;Q_0;-P^0/(\Lambda_2\Lambda_3),
P^0/(\Lambda_1\Lambda_3),-P^0/(\Lambda_1\Lambda_2)\Big)\label{cvecthree}\\
  \Gamma_{IV} &=&
{1\over 4}  \Big(P^0;P^0/\Lambda_1,P^0/\Lambda_2,-P^0/\Lambda_3;Q_0;-P^0/(\Lambda_2\Lambda_3),
-P^0/(\Lambda_1\Lambda_3),P^0/(\Lambda_1\Lambda_2)\Big) \label{cvecfour}
 \end{eqnarray}
in a notation where the $8$ entries of the charge vectors are those of $D6$, three kinds of $D4$'s, $D0$, and 
three kinds of $D2$'s. The total charge vector 
$$
\Gamma =  \Gamma_{I}  + \Gamma_{II}+  \Gamma_{III} + \Gamma_{IV} =(P^0;\vec{0}; Q_0; \vec{0})
$$ 
is that of the $D0-D6$ black hole, as it should be. The direct derivation of (\ref{cvecone}-\ref{cvecfour}) 
can be carried out using formulae in \cite{GLS07} but we will also verify these expressions in section
5.1 of the present paper, using a simple duality chain. 

To interpret the expressions (\ref{cvecone}-\ref{cvecfour}) we compare with a microscopic model based on 
coincident $D$-branes. In the absence of external $B$-fields it has long be known \cite{Taylor:1997ay,Larsen:1999pu} 
that a total charge vector with just $D0$- and $D6$-brane charge can be reproduced using 
four $D6$'s wrapping $T^2\times T^2\times T^2$ with flux assignments :
\begin{align}
  (F_{12},\,F_{34},\,F_{56})^I &= (f_1,\,f_2,\,f_3)~,  \label{eq:fluxone}\\
  (F_{12},\,F_{34},\,F_{56})^{II} &= (f_1,\,-f_2,\,-f_3)~,
   \label{eq:fluxtwo}\\ (F_{12},\,F_{34},\,F_{56})^{III} &=
  (-f_1,\,f_2,\,-f_3)~,  \label{eq:fluxthree}\\
  (F_{12},\,F_{34},\,F_{56})^{IV} &= (-f_1,-\,f_2,\,f_3)\,.
 \label{eq:fluxfour}
\end{align}
The superindex $\{I,\,II,\,III,\,IV\}$ enumerates the four $D6$'s while the subindex 
in the individual fluxes $f_i$ refers to the torus in which they are thread. 
The induced $D0$-brane charge from the flux $F$ is the third Chern class \footnote{The B-field contributions to the 
WZW term should not be included here. Our definition of charges from the four dimensional fields 
(\ref{eq:magfields}) gives nice quantization rules and transformation properties under $U$-duality 
but it does not include the contribution from $B$-fields.}
\begin{equation}
n_0 = -\frac{n_6}4\,{1\over 6(2\pi)^3}\int {\rm tr} F\wedge F \wedge F = -{n_6V_6
f_1f_2f_3\over (2\pi)^3}~,
\label{eq:inddzero}
\end{equation}
so we have
\begin{equation}
{P^0\over Q_0} = {M_6\over M_0}{n_6\over n_0} = -{V_6\over
(2\pi)^6\alpha^{\prime 3}} {(2\pi)^3\over V_6 f_1f_2f_3}
 =  - {1\over (2\pi\alpha^\prime)^3f_1f_2f_3}~.
 \end{equation}
In order to induce the correct total $D0$-brane charge the fluxes must be chosen so that
\begin{equation}
(2\pi\alpha^\prime)^3 f_1f_2f_3  = - {Q_0\over P^0}~.
\label{eq:prodflux}
\end{equation}
The pattern of signs in the fluxes (\ref{eq:fluxone}-\ref{eq:fluxfour}) were chosen such that all induced 
$D2$-brane and $D4$-brane charges cancel. The four $D6$-branes with fluxes arranged in the 
manner indicated are mutually local and they have the total quantum numbers expected of the primitive 
constituents underlying the $D0-D6$ black hole. However, the model is incomplete because it only 
specifies the product of the fluxes $f_1, f_2, f_3$, not their individual values. 

The constituent charge vectors (\ref{cvecone}-\ref{cvecfour}) determined in our construction are
precisely those of $D6$-branes with fluxes (\ref{eq:fluxone}-\ref{eq:fluxfour}) if we identify
the parameters $\Lambda_i$ with fluxes according to
\be
\label{matchf}
\Lambda_i =  - {1\over 2\pi\alpha' f_i}~.
\ee
As a consistency check we note that the constraint (\ref{eq:prodlambda}) maps to
(\ref{eq:prodflux}) under the identifications. More importantly, the constraints 
(\ref{eq:prodlambda}-\ref{eq:sugcons}) determine the $\Lambda_i$'s completely in terms
of the charges $Q_0, P^0$ and the $B$-fields, $B_i$. The identifications (\ref{matchf})
therefore specify the fluxes on the constituent  $D6$-branes completely.

We now have the ingredients to discuss the mass of the $D0-D6$ black hole in a 
background with three independent $B$-fields. The Dirac/Born-Infeld (DBI) mass of the $D6$ 
branes with flux is
\begin{equation}
M = T_6 \int {\rm Tr} \sqrt {{\rm det} \[ G+ (2\pi\alpha^\prime F- B)\]}~.
\end{equation}
We are not using this action in a truly non-abelian setting: we have in mind a
diagonal configuration describing four branes, each of which is BPS by itself
although they are not mutually BPS (they preserve different supersymmetries).
Since $F$ is a rank four bundle, we need to consider $n_6/4$ such objects to 
obtain the correct $D6$-brane charge. This gives a block diagonal configuration 
with unit metric, three B-fields, and the fluxes arranged as above. The mass 
becomes
\begin{eqnarray}
\label{DBI}
M &=& T_6 V_6 \left[ ( 1+ (2\pi\alpha^\prime f_1 - B_{1})^2) ( 1+
(2\pi\alpha^\prime f_2 - B_{2})^2) ( 1+ (2\pi\alpha^\prime f_3 -
B_{3})^2)\right]^{1/2} \cr
 & +&T_6 V_6 \left[ ( 1+ (2\pi\alpha^\prime f_1 - B_{1})^2)
( 1+ (2\pi\alpha^\prime f_2 + B_{2})^2) ( 1+ (2\pi\alpha^\prime f_3 +
B_{3})^2)\right]^{1/2} \cr & +&T_6 V_6 \left[ ( 1+ (2\pi\alpha^\prime f_1 +
B_{1})^2) ( 1+ (2\pi\alpha^\prime f_2 - B_{2})^2) ( 1+ (2\pi\alpha^\prime f_3
+ B_{3})^2)\right]^{1/2} \cr &+&T_6 V_6 \left[ ( 1+ (2\pi\alpha^\prime f_1 +
B_{1})^2) ( 1 + (2\pi\alpha^\prime f_2 + B_{2})^2) ( 1+ (2\pi\alpha^\prime
f_3 - B_{3})^2)\right]^{1/2}~.\label{eq:massformula}
\end{eqnarray}
The total mass depends on the B-fields both {\em explicitly}
as they appear in eq.(\ref{DBI}) and {\em implicitly}, as they determine the proper 
choice of $f_i$'s. The general $D0-D6$ mass formula (\ref{eq:massformula}) computed 
from the DBI formula agrees with the one found (in (5.56) of \cite{GLS07}) by U-duality 
from our seed solution. 

\subsection{Quantization of Charges and Fluxes}
So far we have treated fluxes and charges as continuous variables. This is reasonable
for most purposes, since the objects we study are large. However, the quantization 
conditions lead to several important refinements which we turn to next. 

One aspect of quantization is due to our constituent model having exactly four 
primitive constituents. For the $D0-D6$ black hole these four constituents appear on an 
equal footing in that they can be permuted by symmetries. This structure is 
consistent with the underlying charge quantization if the underlying numbers of 
$D0$- and $D6$-branes both are divisible by four, but otherwise not. Non-BPS states 
with charges that are not divisible by four are therefore protected against spontaneous
separation into primitive constituents. Generally there will be a binding energy that is
finite, although not parametrically large. In other words, the binding energy will be 
microscopic even for macroscopic states, leaving the state fragile rather than unbound . 
This binding mechanism is a non-BPS version of the customary restriction to mutually prime quantum 
numbers for threshold BPS bound states. 

Another aspect of the quantization conditions concerns the flat directions in moduli space. 
Recall that for non-BPS black holes there are two moduli that fail to be stabilized by the attractor
mechanism, even though they are in vector multiplets. In the $D0-D6$ frame these unfixed 
moduli are just ratios of the three torus volumes, $v_1,v_2,v_3$. The attractor mechanism does apply
to the overall volume torus $V_6$ and the three $B$-field densities, so these should be kept fixed as
we move along the flat directions. It is immediately apparent that the dimensionful constituent charge 
vectors (\ref{cvecone}-\ref{cvecfour}) are {\it independent} of the flat directions. 

Now, the relation between the dimensionful constituent charge vectors and the corresponding 
quantized charges depends on moduli. For example 
the dimensionless vector corresponding to (\ref{cvecone}) becomes:
\begin{equation}
\label{quantvector}
  \vec{n}_{I} ={1\over 4}
  \Big(n_6;-n_6/\lambda_1,-n_6/\lambda_2,-n_6/\lambda_3; n_0; n_6/(\lambda_2\lambda_3),
  n_6/(\lambda_1\lambda_3),n_6/(\lambda_1\lambda_2)\Big)\,,
 \end{equation}
where $\lambda_i = \Lambda_i/v_i$. The flux densities $\Lambda_i^{-1}$ 
remain invariant under volume changes but the scaled densities $\lambda_i^{-1}$ vary. 
The charge split for a non-BPS bound state with the same charges as $n_6$ 
$D6$-branes and $n_0$ $D0$-branes therefore varies as we move along the flat direction.
Moreover, the number of $D4$-branes and $D2$-branes of the primitive constituents depend
continuously on the flat moduli. The lack of proper quantization implies some
some interesting finite $g_s$ corrections to our picture of the non-BPS extremal black hole 
as a marginal bound state, but we will not develop this point further in this paper. 

\subsection{Multi-center BPS solutions and decay of the non-BPS $D0-D6$ black holes}
Our extremal single-center non-BPS black hole solutions have the same quantum 
numbers as a certain class of multi-center BPS solutions. Some of the characteristic features 
of these BPS multi-center solutions are :
\begin{itemize}
\item Their mass is BPS, which is always {\it strictly} smaller than that of  the non-BPS solution with otherwise
identical 
quantum numbers (see \cite{GLS07}).
\item
They are bound states of as few as two 1/2-BPS constituents. (We interpret non-BPS 
solutions in terms of exactly four 1/2-BPS constituents.)
\item
The charge vectors of the 1/2-BPS constituents are mutually non-local, {\it i.e.} they have non-zero 
intersection number. (The four constituents of the non-BPS black holes are mutually local.) 
\item
The constituents have a finite separation scale that is essentially determined by the charge intersection
numbers. (The constituents of the non-BPS black holes can move freely in the 
supergravity approximation.)
\item 
These BPS states only exist in part of the moduli space. There is a 
co-dimension one wall of threshold stability in moduli space beyond which they
disappear from the spectrum. (The non-BPS black holes exist everywhere in moduli space.) 
\item
The mutual non-locality of the charges generally necessitates angular momentum in the multi-center
BPS solutions. Varying the location of the constituents gives a range of allowed
angular momenta identical to the so-called ``slowly'' spinning non-BPS black hole.
\end{itemize}
The multi-center BPS solutions are thus very different from the non-BPS solutions analyzed
in this article. For example, it is evident that the BPS solutions cannot be continuously connected to 
any non-BPS stationary solution through the wall of marginal stability, as illustrated in the figure below. Instead, there can be decay from the 
non-BPS branch to the BPS branch on the part of moduli space where BPS solutions exist. 
The transition will release energy, entropy and generally also angular momentum. This indicates a first order 
transition between the two branches.  

The $D0-D6$ duality frame is the simplest setting for making this discussion more explicit. Then the 
two types of constituents on the BPS branch are just $D0$-branes and $D6$-branes\footnote{The BPS branch may have other components corresponding to solutions with more than 
two types of constituents or with the same number of different constituents. We do
not explore this possibility here.}.  We will briefly 
summarize some of the history of BPS $D0-D6$ bound states.

The problem of adhering $D0$-branes to $D6$-branes in a supersymmetric manner was 
first considered by Witten \cite{Witten:2000mf}\footnote{See also related work in \cite{Mihailescu:2000dn}.}. He found that a supersymmetric branch 
exists for sufficiently large $B$-fields
\begin{equation}
\label{eq:boundary}
\sum_{i<j} B^i B^j \geq 1~.
\end{equation}
The equality defines a wall of threshold stability, which in the equal B field case is represented as the dashed line in the figure below. 

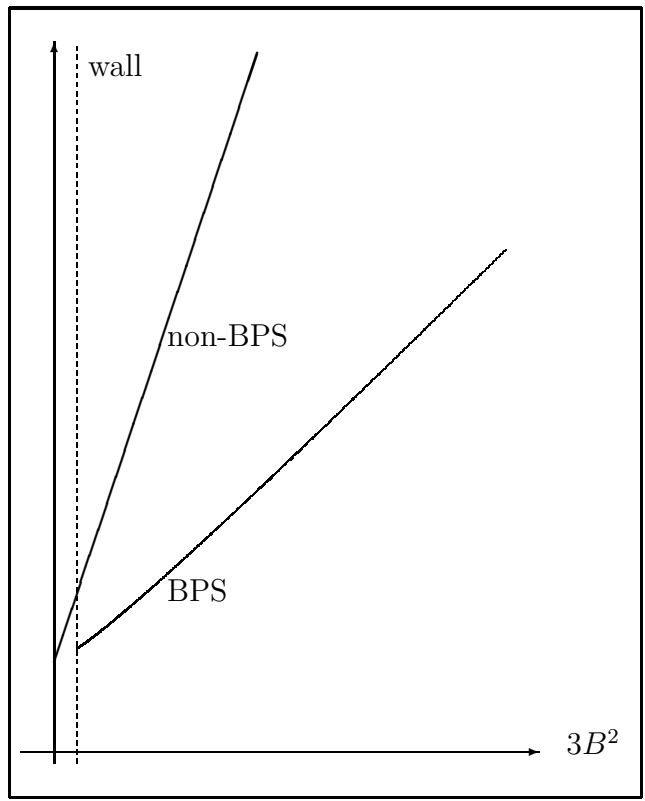
\begin{figure}
\begin{center}
\setlength{\unitlength}{0.3cm}
\begin{picture}(24,35)(-2,-2)
\put(-1.5,0){\vector(1,0){23}}
\put(22.7,-0.1){$3B^2$}
\put(0,-0.5){\vector(0,1){32}}
\qbezier(1,4.58258)(4.2423,6.7052)(20,22.2711)
\multiput(1,-0.5)(0,0.4){80}{\line(0,1){0.2}}
\put(1.5,30){wall}
\put(5, 6.7){BPS}
\thicklines
\put(0,4){\line(1,3){9}}
\put(5,18){non-BPS}
\put(-2,-2){\line(1,0){28}}
\put(-2,33){\line(1,0){28}}
\put(-2,-2){\line(0,1){35}}
\put(26,-2){\line(0,1){35}}
\end{picture}
\end{center}
\caption{Mass of the non-BPS and BPS branches for diagonal B field}
\end{figure}

In our notation we write the condition as 
$\sin\alpha < 0 $ where $\alpha$ is the phase of the spacetime central charge matrix
\be
e^{i\alpha} \doteq \frac{Q_0 + iP^0 \prod_{i=1}^3 (1 + i B_{i})}{\big|Q_0 + iP^0 \prod_{i=1}^3 (1 + i B_{i}) \big|}~.
\ee
(We take $P^0, Q_0>0$ without loss of generality.) 
Near the boundary $\alpha=0$, open strings stretching between the $D0$-branes 
and the $D6$-branes have light modes described by an effective quantum mechanics. 
The supersymmetric branch $\alpha<0$ is the Higgs phase of this theory. Here
there is a tachyonic open string mode which condenses to a supersymmetric ground state, interpreted
as the $D0-D6$ bound state. This BPS state disappears from the spectrum upon crossing the wall to $\alpha>0$.


In the early work no BPS supergravity solution was detailed. Later progress in the field uncovered
explicit supergravity BPS multi-center solutions \cite{Denef:2000nb,Denef:2002ru} which carry $D0-D6$ 
charge. These are representations of the perturbative bound state in the supergravity regime.  
We are particularly interested in the two-center solution based on two 1/2-BPS charge vectors,
\begin{eqnarray}
\Gamma_1 &= &(P^0,0,0,0)~,\cr \Gamma_2 & = &(0,0,0,Q_0)~,
\end{eqnarray}
which are not mutually local,
\begin{equation}
\langle \Gamma_1 , \Gamma_2 \rangle = P^0 Q_0~,
\end{equation}
but they sum up to the total $D0-D6$ charge vector $\Gamma$.  Following \cite{Denef:2002ru}
one can show that a two-center BPS solution\footnote{For recent work on supersymmetric D0-D6 supergravity configurations, see \cite{Lee:2008ha,Castro:2009ac}.} with these charge assignments exists exactly when the separation between the two centers is :
\begin{equation}
\label{constraint}
R = |\vec{x}_1-\vec{x}_2| = - \frac{P^0}{ \sin\alpha}
=  { |Q_0 + iP^0\prod_{i=1}^3 (1 + i B^i)|
\over \sum_{i<j} B^i B^j -1}~.
\end{equation}
The supergravity scale (\ref{constraint}) is meaningful only for $\sin\alpha < 0$, 
exactly the same as the condition for supersymmetry in the analysis of light open string 
modes.

The agreement of the BPS moduli space for the supergravity solution and the perturbative analysis 
can be understood as follows \cite{Denef:2002ru}. There are {\it three} regimes in parameter 
space where the $D0-D6$ bound state can be simply analyzed: a ``Higgs'' branch, a
``Coulomb'' branch, and the multi-center supergravity solution. At any point on BPS moduli 
space, the scale (\ref{constraint}) is proportional to the charges $Q_0, P^0$, which 
in terms are proportional to the coupling $g_s$ (multiplied by the number of branes). For 
vanishing string coupling the scale is negligible so we can focus on the lightest open string modes 
which, as mentioned earlier, are tachyonic. Condensation of these modes fixes the separation scale 
at exactly zero. This is the Higgs branch. Any non-vanishing scale \eqref{constraint} renders the open string modes massive.  This occurs whenever the string coupling is non-vanishing. As the scale becomes larger, the Higgs potential becomes more shallow, and the semiclassical approximation for the Higgs branch breaks down because the wave function spreads out more. It is the integration of these massive modes out that leads to an effective potential for the $D0-D6$ separation and the scale (\ref{constraint}) is a consistent minimum of that potential. This is the Coulomb branch description. 
Finally, if the separation scale fixed by the minimum of the potential lies 
beyond the string scale, a supergravity analysis becomes more reliable than the quantum mechanics 
derived from looking at just the low-energy open string modes. In summary, increasing the string coupling 
$g_s$ from zero pushes us through a cascade of useful regimes: ``Higgs''  to ``Coulomb'' to supergravity.
Since this is true anywhere on the BPS moduli space $\sin\alpha<0$ it follows that this space comes out 
the same at weak and strong coupling. 

We can also use the scale (\ref{constraint}) as a guide towards the physics near the boundary 
of supersymmetric moduli space. For any finite value of $g_s$ the scale increases without bound 
as $\alpha\to 0^-$. Even if the fixed value of $g_s$ is so small that the Higgs description applies 
initially, the motion towards the boundary $\alpha=0$ will again force us through the cascade 
of useful descriptions, from ``Higgs''  to ``Coulomb'' to supergravity. For example, the Higgs branch
analysis becomes unreliable as the variance in the expectation value for the tachyonic string modes 
becomes large.  Once we reach a multi-center configuration, the separation between the $D0$-branes 
and the $D6$-branes diverges as we take $\alpha \to 0^-$.  This justifies our contention that we are 
dealing with a threshold transition where BPS states completely exit from the spectrum.

The significance of this discussion for the non-BPS supergravity solutions we focus on in this paper 
is primarily that it falsifies an alternative hypothesis: the wall of marginal stability does {\it not} 
indicate a continuous transition to a non-BPS branch. Such a transition is also excluded on the grounds
that the non-BPS branch exists for all moduli. Moreover, for the moduli $\sin\alpha<0 $ where 
both branches exist, the non-BPS states have larger energy, so there we expect a first order transition 
between the branches. 

Angular momentum provides an important elaboration to this picture. The simple two-center 
$D0-D6$ BPS solution discussed above is supported by angular momentum $|\vec{J}| = {1\over 2}P^0Q_0$. 
A more general family of solutions with the same total charge vector $\Gamma$ has the $D6$ at the 
center and places the $D0$'s on a sphere of the radius $R$ given in (\ref{constraint})
\cite{Denef:2002ru}. The total angular momentum of this "halo" solution is weigted by the $D6$ distribution 
such that it covers the entire range $0 \le |\vec{J}| \le {1\over 2}P^0Q_0$. This is the same range of 
allowed angular momenta found for the ``slowly rotating" extremal non-BPS black 
hole \cite{Larsen:1999pp}. The first order transition from the non-BPS to the BPS branch can therefore 
proceed for any of these allowed angular momenta, and for the entire range of moduli with $\sin\alpha < 0 $.

There are other similarities between the two branches. For example, 
it is interesting that the multi-center BPS solutions share the same flat directions as 
their extremal non-BPS black hole cousins. At first this may seem to contradict our previous 
statement about BPS black holes being complete attractors. However, the multi-center solutions
do not have regular horizons, they have primitive 1/2-BPS components with singular behavior at 
their "horizons".  This structure is precisely what is needed to allow an orbit of multi-center 
solutions under the same non-compact group as for the extremal non-BPS black hole.

\section{Non-BPS Black Holes from $D3$-branes at Angles}
In order to be sufficiently general, our generating ${\overline D0}-D4$ solution must allow for
an ambient $B$-field, and the dual $D0-D6$ black holes must have $3$ independent $B$-fields. There
is yet another useful duality frame, where all the charges of the black holes are those of $D3$-branes. 
In this frame the inevitable additional parameter is incorporated geometrically, as a relative angle 
of the intersecting $D3$-branes. 

As an application of the $D3$-brane representation we discuss the perturbative open string 
analysis of the intersecting $D3$-brane system. We also discuss the supersymmetry preserved
by various subsets of $D3$-branes, but broken by the system as a whole. 

\subsection{Duality from $D0-D6$ to $D3$-branes at Angles}
As we have emphasized, it is advantageous to analyze the $D0-D6$ system in terms of its
four primitive constituents with charge vectors (\ref{cvecone}-\ref{cvecfour}).
These are four constituent $D6$-branes with fluxes, which we turn
into four constituent $D3$-branes by acting with T-duality three times, once along each of the 
three $T^2$'s. On each $T^2$, T-duality along one direction turns the flux into 
the angle that the resulting brane subtends with respect to the transverse direction.

To be more precise, the world-volume theory of a $D$-brane
is determined entirely by the gauge invariant quantities
\begin{equation}
2\pi\alpha^\prime{\cal F}_i 
= 2\pi\alpha^\prime\,f_i - B_i = - (\Lambda^{-1}_i + B_i)~,
\label{eq:gaugeinv}
\end{equation}
where we used the notation (\ref{matchf}) for the inverse fluxes. The resulting angle
with respect to the undualized direction on the $i$th $T^2$ is then determined by
\begin{equation}
   \cot \phi_i = 2\pi\alpha^\prime{\cal F}_i = - (\Lambda^{-1}_i + B_i)~.
 \label{eq:tdual}
\end{equation}

In our set-up the T-duality acts on each of the four constituent branes enumerated by $A, B= I, II, III, IV$
and endowed with the fluxes (\ref{eq:fluxone}-\ref{eq:fluxfour}). For each pair of branes the
fluxes are the same on one of the three $T^2$'s and it has the opposite sign on the remaining $T^2$'s.
If the fluxes agree, the branes are obviously parallel within that $T^2$, {\it i.e.} their
relative angle is $\vartheta^{AB}_i = \phi^A_i - \phi^B_i = 0$. If the fluxes have opposite signs,
the relative angle of the brane pair within that torus is
\begin{equation}
 \cot \vartheta^{AB}_i  = \cot(\phi^A_i - \phi^B_i) = \frac{1 -
\cot(\phi^A_i)\cot(\phi^B_i)}{\cot(\phi^A_i) + \cot(\phi^B_i)}
 =  - \frac{1}{2}\, [\Lambda_i(1 + B_i^2) -
  \Lambda_i^{-1}] ~.\label{fluxes}
\end{equation}
Here $\Lambda_i$ refers to brane $A$, in contrast to $\Lambda_i^B = -\Lambda_i$.
At this point we recall that the parameters $\Lambda_i$ are determined by charges and $B$-fields 
according to (\ref{eq:prodlambda}-\ref{eq:sugcons}). In fact the constraints (\ref{eq:sugcons})
demand that the expression on the right hand side of (\ref{fluxes}) is the same for each of the $T^2$'s. 
\begin{equation}
\cot \vartheta^{AB}_i=  - \frac{1}{2}\, [\Lambda_i(1 + B_i^2) -\Lambda_i^{-1}] \equiv -b~.
\label{eq:relangle}
\end{equation}
In summary, after duality each pair of constituent $D3$-branes have the same relative angle (\ref{eq:relangle})
within two of the $T^2$'s, and they are parallel within the last $T^2$. 

In the preceding formulae we have been sloppy with signs, in order to keep notation simple: in 
\eqref{eq:gaugeinv} we took the field strength $f_i$ on the $i$th $T^2$. According to the flux assignments
(\ref{eq:fluxone}-\ref{eq:fluxfour}) this is accurate for the constituent $A=I$ but for $A=II, III, IV$ some of 
the fluxes actually have their signs flipped. The overall sign in our final formula for the relative angles
\eqref{eq:relangle} should be adjusted accordingly.

The constant $b$ has a simple interpretation, mentioned already after (\ref{eq:sugcons}). It is the $b$-field
in the $\overline{D0}-D4$ seed solution that gave rise to the $D0-D6$ black hole after dualities. It now 
appears in the expressions for the angles subtended by $D3$-branes. We can take the dualities full 
cicle by aligning our coordinate system with one of the $D3$-branes, and then dualize it to a $D0$ brane. 
Under this duality the remaining $D3$-branes turn into $D4$-branes. This sequence of dualities can be viewed
as an independent derivation of the $D0-D6$ primitive constituents (\ref{cvecone}-\ref{cvecfour}). In particular, 
it gives confidence in the nonlinear constraints (\ref{eq:sugcons}) imposed on the fluxes in the $D0-D6$-frame.


\subsection{Perturbative Stability of Extremal Black Holes}
We can study the stability of the extremal non-BPS black holes at weak
string coupling by examining the open strings stretching between its four constituent
$D$-branes. We will do this in the $D3$-brane duality frame where the fluxes 
are encoded in the angles between the branes. The corresponding discussion for
the $D0-D6$ system (and other frames) then follow by duality \footnote{The duality is
from $D3$-branes at angles to the {\it constituents} of the $D0-D6$-system. We can 
also apply formulae similar to the ones below directly to the $D0-D6$ branes 
(following \cite{Witten:2000mf}) but in that case all three angles are turned on, giving rise
to tachyons and spacetime supersymmetry breaking.}. 

We first consider a single pair of $D3$-branes on $T^6$. Each of the $D3$'s have one 
direction on each of the three $T^2$, and their relative angles on the three $T^2$'s are 
$\theta_i$ with $0 \le \theta_i \le \pi$. The boundary conditions on the open 
strings stretching between the $D3$'s are then twisted due to the relative angles.
For example, the complex scalars on the three $T^2$ have fractional modes 
$X^i_{n+\theta_i/\pi}$ and fermion modes are similarly twisted. Summing up the twisted 
ground states energies and keeping only the GSO projected states, the lightest states 
in the NS-sector become four complex scalars in spacetime with masses 
(for more details see {\it e.g.} \cite{rabadan}):
\begin{align}
\alpha^\prime   m_1^2 &= \frac{1}{2\pi}\left(-\theta_1+\theta_2 + \theta_3\right)\,, \nonumber \\
\alpha^\prime   m_2^2 &= \frac{1}{2\pi}\left(\theta_1-\theta_2 + \theta_3\right)\,, \nonumber \\
\alpha^\prime   m_3^2 &= \frac{1}{2\pi}\left(\theta_1+\theta_2 - \theta_3\right)\,, \nonumber \\
\alpha^\prime   m_4^2 &= 1 -\frac{1}{2\pi}\left(\theta_1+\theta_2 + \theta_3\right)\,.
 \label{eq:masses}
\end{align}
It is clear that for some values of the relative angles the perturbative spectrum contains tachyons,
interpreted as a classical instability. 
A pair of constituent $D3$-branes is stable if their relative angles satisfy the {\it stability conditions} :
\begin{align}
  \theta_2 + \theta_3 &\geq \theta_1\,, \nonumber \\
  \theta_1 + \theta_3 &\geq \theta_2\,, \nonumber \\
  \theta_1 + \theta_2 &\geq \theta_3\,, \nonumber \\
  2\pi & \geq  \theta_1 + \theta_2 + \theta_3\,,
\label{eq:stability}
\end{align}
which assure the absence of tachyons. The non-BPS extremal black hole is perturbatively
stable if these conditions are satisfied for any pair of constituents.

The stability conditions are generally quite complicated. However, as we summarized after 
\eqref{eq:relangle}, the relevant angles are very simple in our case: for each pair of branes one 
of the relative angles vanish, while the other two angles have identical norm which we can 
choose in the range $0\leq |\vartheta^{AB}_i| \leq\pi$. Identifying $\theta_i=|\vartheta^{AB}_i|$, 
the stability conditions \eqref{eq:stability} are easily verified. 
In more detail, the spectrum of light open strings \eqref{eq:masses} takes the values
\begin{equation}
\alpha^\prime m^2 = 0~, 0~, \frac{\theta}{\pi}~, 1-\frac{\theta}{\pi}~,
\label{eq:bspec}
\end{equation}
for each pair of branes, where $\theta$ is the only non-trivial angle with $0\leq\theta\leq\pi$. 
These masses are non-negative.

In the R-sector the ground state energy vanishes by world-sheet supersymmetry. The 
four complex fermion oscillators $\psi^i_{n+\theta_i/\pi}$ (with $i=0,1,2,3$) then have 
positive frequency modes $0, 0, \theta/\pi, 1-\theta/\pi$. The GSO projected states
have exactly one fermion oscillator so we recover the spectrum \eqref{eq:bspec},
in accordance with spacetime supersymmetry satisfied by each pair of $D3$-branes.

In the $D3$-frame we have seen that the equality of the relative angles on two different tori 
gives a simple solution to the stability conditions \eqref{eq:stability} for each pair. The stability 
conditions are less transparent in other duality frames. In the $D0-D6$ frame, 
the dictionary \eqref{fluxes} identifies the relative angles with the obscure combinations 
$-{1\over 2}[\Lambda_i (1+B_i^2) - \Lambda^{-1}_i]$. The requirement
\eqref{eq:sugcons} that these combinations be the same on the three $T^2$'s can 
therefore be interpreted as a stability condition in the $D0-D6$ frame. 

Considering the second equation in \eqref{eq:relangle} we can alternatively identify the relative 
angles of the $D3$-branes on the three $T^2$'s with the background $B$-fields in 
the $\overline{D0}-D4$ seed solution. The absence of open string tachyons among pairs of the 
$\overline{D0}-D4$ branes is thus upheld by the choice of a diagonal $B$-field in this frame.

The perturbative open string analysis evidently focusses on pairs of 
constituent $D$-branes. However, as we detail in the following section, the total breaking of 
supersymmetry can be seen only when all four 1/2-BPS constituents are taken into account. 
Since each of the pairs preserves some supersymmetry, the absence of tachyons from the 
open string spectrum was therefore anticipated on general grounds. 
The complete spectrum of the non-BPS black hole includes collective states that depend
for their existence on the presence of three and four branes. Only the last kind, depending 
on all four constituent branes, are sensitive to supersymmetry breaking. Since the 
classical non-BPS black hole entropy vanishes unless all four constituent branes are 
present, we expect numerous modes of this type. The finite entropy and sensible 
thermodynamics do not suggest any instability among these more exotic modes so we 
expect any tachyons in this sector either.

The instability to the BPS branch discussed in section 4.3 is not manifested as a tachyon 
even among the modes depending on the presence of four branes, because this instability is 
nonlocal in character: the non-BPS states can be represented geometrically as four $D3$-branes intersecting over 
complex lines, while the corresponding BPS states are just two $D3$-branes intersecting over a 
single point. In the region of moduli space where the BPS bound state exists, both 3-surfaces have 
locally minimal areas, so transition from one type of surface to the other will
be non-perturbative. This is consistent with the first order phase transition we discussed in
section 4.3.

\subsection{Supersymmetry Breaking of $D3$-branes at Angles}
The black holes we study are non-BPS but their primitive constituents are $1/2$-BPS. It is interesting 
to examine how supersymmetry is broken by the addition of the different components of the bound 
state. Our discussion will be in the $D3$ duality frame for definiteness but the supersymmetry structure
is the same for the four primitive constituents in other frames. 

In order to compare the preserved supercharges of one $D3$-brane with those preserved by 
another, we must rotate the supercharges. The generators of rotations within the $T^2$'s are :
 \be
S_1 = \frac{i}2\, \Gamma^1\Gamma^2~, \qquad S_2 = \frac{i}2\, \Gamma^3\Gamma^4~,
\qquad S_3 = \frac{i}2\, \Gamma^5\Gamma^6~.
 \ee
It is useful to organize the preserved supercharges into eigenvectors
of the rotation generators \cite{micha,vijay-rob} :
\beq
 S_i | s_1~,s_2~,s_3\rangle &=&\frac{1}{2}s_i \,| s_1~,s_2~,s_3\rangle~, \nonumber
 \eeq
where $s_i=\pm$. Each supercharge $|s_1~,s_2~,s_3\rangle$ is also a chiral spinor in the two 
non-compact dimensions, and so corresponds to two supercharges. Thus a single $D3$-brane 
preserves $16$ supercharges, $1/2$ of the maximal supersymmetry. 

Consider a pair of constituent $D3$-branes of type $A, B$ situated at the relative angles 
$\vartheta^{AB}_1, \vartheta^{AB}_2, \vartheta^{AB}_3$ within the three $T^2$'s. A candidate 
supersymmetry $|s_1~,s_2~,s_3\rangle$ preserved by one of these is preserved also by the other when
\be
s_1 \vartheta_1^{AB}  + s_2 \vartheta_2^{AB} + s_3 \vartheta_3^{AB} \;=\; 0\;
\text{mod}\; 2\pi~.
\label{eq:susycond}
\ee
When applying these conditions we must be careful with the {\em sign} of the angle $\vartheta_i^{AB}$.
We will follow the definition \eqref{eq:relangle} when determining relative signs. 

Consider for example the constituents $I$ and $II$. The fluxes (\ref{eq:fluxone}-\ref{eq:fluxtwo})
mean the corresponding $D3$-branes are aligned on the first $T^2$ so $\vartheta^{I,II}_1=0$. On the
remaining $T^2$'s the $D3$-branes generally meet at a nontrivial angle. In fact they meet at the 
{\it same} relative angle $\vartheta^{I,II}_2=\vartheta^{I,II}_3$ on the remaining $T^2$'s. The supersymmetry
condition \eqref{eq:susycond} then correlates the spinorial indices of the supersymmetries on those last $T^2$'s:
both $D3$-branes preserve $|s_1~, +~, - \rangle$ and $|s_1- +\rangle$ for $s_1=\pm$. 
Proceeding similarly for the others pairs of the form $(I, B)$ we find the results reported in table \ref{tab:pairIB}.

\begin{table}[h!]
  \centering
  \setlength{\extrarowheight}{3pt}
  \renewcommand{\arraystretch}{1.3}
    \begin{tabular}{|>{$}l<{$}|>{$}l<{$}|>{$}l<{$}|}\hline
      \multicolumn{1}{|c|}{Pair} &   \multicolumn{1}{c|}{Alignment angle} &
      \multicolumn{1}{c|}{Supersymmetry} \\
      \hline\hline
      (I,\,II) & \vartheta^{I,II}_1=0 & |s_1~,-~,+\rangle~, |s_1~,+~,-\rangle \\[3pt]
       (I,\,III) & \vartheta^{I,III}_2=0 & |-~,s_2~,+\rangle~, |+~,s_2~,-\rangle \\[3pt]
       (I,\,IV) & \vartheta^{I,IV}_3=0 & |-~,+~,s_3 \rangle~, |+~,-~,s_3 \rangle \\[3pt]
      \hline
    \end{tabular}
  \vspace{8pt}
  \caption{$(I,\,B)$ pairs of constituents and their preserved supersymmetries.}
  \label{tab:pairIB}
\end{table}

The remaining pairs of $D3$-branes have the opposite relative orientation. 
For example, consider the constituents $III$ and $IV$. The 
fluxes (\ref{eq:fluxthree}-\ref{eq:fluxfour}) again correspond to $D3$-branes that are aligned in first $T^2$ so 
$\vartheta^{III,IV}_1=0$. However, they meet at the {\it opposite} relative angles 
$\vartheta^{III,IV}_2=-\vartheta^{III,IV}_3$ on the 
remaining $T^2$'s. The supersymmetry condition \eqref{eq:susycond} again imposes no condition on the 
first $T^2$ but now it {\it anti}-correlates the remaining $T^2$'s, leaving the preserved supersymmetries 
$|s_1~, -~,- \rangle$ and $|s_1~,+~,+\rangle$ 
for $s_1=\pm$. Proceeding similarly for the remaining pairs we find the results reported in table \ref{tab:pairBC}.

\begin{table}[h!]
  \centering
  \setlength{\extrarowheight}{3pt}
  \renewcommand{\arraystretch}{1.3}
    \begin{tabular}{|>{$}l<{$}|>{$}l<{$}|>{$}l<{$}|}\hline
      \multicolumn{1}{|c|}{Pair} &   \multicolumn{1}{c|}{Alignment angle} &
      \multicolumn{1}{c|}{Supersymmetry} \\
      \hline\hline
      (III,\,IV) & \vartheta^{III,IV}_1=0 & |s_1~,-~,-\rangle ~, |s_1~,+~,+\rangle \\[3pt]
       (II,\,IV) & \vartheta^{II,IV}_2=0 & |-~,s_2~,-\rangle ~, |+~,s_2~,+\rangle \\[3pt]
       (II,\,III) & \vartheta^{II,III}_3=0 & |-~,-~,s_2 \rangle~, |+~,+~,s_3 \rangle \\[3pt]
      \hline
    \end{tabular}
  \vspace{8pt}
  \caption{$(B,\,C)$ pairs of constituents and their preserved supersymmetries.}
  \label{tab:pairBC}
\end{table}

Each pair of $D3$-branes preserve $8$ supersymmetries, $1/4$ of maximal supersymmetry. We see
from the tables that any three constituent $D3$-branes still preserve four supersymmetries, $1/8$ 
of maximal supersymmetry. For example the triple $(I, II, III)$ preserves $|-,-,+\rangle$ and $|+,+,-\rangle$.
It is when we add the fourth constituent that supersymmetry will be broken. For example, the 
supersymmetries preserved by the pair $(I,II)$ have none in common with those preserved by the 
pair $(III,IV)$.

Multiple branes intersecting at angles generally preserve supersymmetry exactly when all the branes
are related by $SU(3)$ rotations \cite{micha,vijay-rob,vijay-finn}. In the examples above, the rotation relating 
each pair of branes is indeed $SU(2)$ with respect to an appropriate complex structure. Also, the
rotations relating each triple fit into an $SU(3)$ for some complex structure. The important point is
that the required complex structures are incompatible. For example, the rotations with 
$\vartheta_2^{III,IV}=-\vartheta_3^{III,IV}$ are $U(1)$ rotations with opposite angles, and so they 
combine to an $SU(2)$ rotation with respect to the most obvious complex structure. The rotations with
$\vartheta_2^{I,II}=\vartheta_3^{I,II}$ similarly combine to an $SU(2)$ rotation, but with respect to 
a complex structure that has the opposite orientation on one or the other $T^2$. 

The preceding analysis was for generic rotation angles $\vartheta^{AB}_i$. Alternatively we can try to
satisfy the supersymmetry conditions \eqref{eq:susycond} by considering special angles. If for each pair we 
can take one of the angles $\vartheta_i^{AB}$ to vanish and the other two identical, either $0$ or $\pi$, 
then {\it all} the $16$ supersymmetries of a single $D3$-brane are preserved. The 
dictionary \eqref{eq:relangle} shows that these special angles appear in the limit 
$b = \pm \infty$.  Therefore full supersymmetry is restored in our $\overline{D0}-D4-D4-D4$ seed solution 
with a diagonal B-field $B$, in the limit where $B \to \pm\infty$.

 \section{Discussion}
We conclude with a few open questions that we hope to address in future work :
\begin{itemize}
\item
{\bf Quantum corrections:} 
Our constituent model has no binding energy at the classical level, and it has
exactly flat directions that are special to the non-BPS branch. It is interesting 
to ask whether these properties are preserved by quantum corrections. 

One indication that they are not is the apparent failure (discussed in section 4.2) of charge 
quantization: the black hole itself carries quantized charges, but it is generally 
not clear why the charges of the individual constituents should be. This is
related to the breaking of the classical $U$-duality group $(SL(2,R))^3$ to its 
discrete version by quantum corrections. This relation may give a way 
to understand the corrections more precisely. 

\item
{\bf The 5D interpretation:} in the absence of $B$-fields the $D0-D6$ solution allows a
a purely geometrical interpretation in $5D$, identified as a near horizon patch
of the extremal Kerr solution \cite{Emparan:2006it}. It would be interesting to extend this identification
to include the three $B$-fields. This will introduce charges in 5D and for this
larger family of solutions there may be limits that are under good control.
\item
{\bf Multi-center solutions:} our arguments suggest the existence of multi-centered 
extremal non-BPS supergravity configurations satisfying two basic requirements : the locations
of the centers should not be constrained and the charge vector at each center should 
should be that of the appropriate constituent model. 

The multi-center non-BPS solutions reported in \cite{Gaiotto:2007ag} seem to confirm
this expectation, for the special case of diagonal charges. It would be interesting to
check this expectation for generic moduli in the more general configurations described
in \cite{Goldstein:2008fq,Bena:2009ev} and to study the differences that will arise in the presence
of angular momentum.

\item
{\bf The microscopic interpretation of non-BPS black hole entropy: } 
classically, the entropy formulae of BPS and non-BPS black holes are almost identical. For example, the 
black holes in $N=8$ supergravity have entropy of the form $S_{\rm BPS}= 2\pi\sqrt{|J_4|}$ with the 
non-BPS/BPS distinction encoded in the sign of the quartic invariant $J_4$. The obvious similarity 
between the entropy formulae on the two branches is commonly interpreted as a hint
that their microscopic origins are virtually identical, {\it i.e} related by analytical 
continuation \cite{Emparan:2006it}. 
Our work challenges this interpretation in its simplest form, by 
highlighting significant differences between the two branches. For example, the 
classical moduli spaces are different even in their dimensionality. This is relevant 
because precise counting of BPS states often involves choosing a favorable point
in moduli space, such as turning on a small $B$-field to avoid bound states at threshold. 
This is not possible for the non-BPS states where the $B$-field is a classical 
modulus. Therefore the corresponding microstates cannot be related by analytical 
continuation. 

Despite these challenges we remain sympathetic to the idea that the extremal non-BPS 
entropy can be understood in a simple manner. The significant differences between the 
two branches must be addressed by a more detailed understanding of the microscopics.
Indeed, they may give guidance towards such a description. 
\end{itemize}

\section*{Acknowledgements}
We thank S. Ferrara, R. Myers, and M. Shigemori for discussions.
JS would like to thank the University of California at Berkeley for
hospitality during part of this work. The work of EG is supported in
part by the US DOE under contract No. DE-AC03-76SF00098 and the
Berkeley Center for Theoretical Physics.  The work of FL is
supported by DoE under grant DE-FG02-95ER40899.
The work of JS was partially supported by the Engineering and 
Physical Sciences Research Council [grant number EP/G007985/1].

\end{document}